\documentstyle[twocolumn,floats,aps,epsf]{revtex}
\newcommand{\k}{{\bf k}}
\def\gtsim{\vbox {\hbox{\lower 0.9\baselineskip \hbox{$>$}} \break
         \hbox{\lower 0.2\baselineskip \hbox{$\sim$}} } }

\begin{document}

\twocolumn[\hsize\textwidth\columnwidth\hsize\csname@twocolumnfalse%
\endcsname

\title{Gap inhomogeneities and the density of states in disordered
d-wave superconductors}
\author{W. A. Atkinson$^1$, P. J. Hirschfeld$^1$, A. H. MacDonald$^2$}
\address{$^1$Department of Physics, University of Florida, PO Box 118440,
Gainesville FL 32611}
\address{$^2$Department of Physics, Indiana University,
Swain Hall W.\ 117, Bloomington IN 47405}
\date{\today}
\maketitle
\draft
\begin{abstract}

We report on a numerical study of disorder effects in 2D $d$-wave BCS
superconductors.  We compare exact numerical solutions of the
Bogoliubov-deGennes (BdG) equations for the density of states
$\rho(E)$ with the standard T-matrix approximation.  Local suppression
of the order parameter near impurity sites, which occurs in
self-consistent solutions of the BdG equations, leads to apparent
power law behavior $\rho(E) \sim |E|^\alpha$ with
non-universal $\alpha$ over an energy scale comparable to the single
impurity resonance energy $\Omega_0$.  We show that the novel effects
arise from static spatial correlations between the order parameter
and the impurity distribution.

\end{abstract}


]
\narrowtext

In spite of strong electronic correlations in the normal
state, the superconducting state of high $T_c$ materials seems
to be accurately described by a conventional BCS-like phenomenology.
The debate over the $k$-space structure of the BCS order parameter
$\Delta_\k$ has now been resolved in favour of pairing states
with $d$-wave symmetry.  Since this symmmetry implies
that $\Delta_\k$ changes sign under rotation by $\pi/2$, there are
necessarily points on the 2D Fermi surface at which $\Delta_\k$
vanishes.  The unique low-energy properties of high $T_c$
superconductors are determined by the quasiparticle  excitations
in the vicinity of these nodal points.

For conventional $s$-wave superconductors, the density of states (DOS)
$\rho(E)$ has a well defined gap and is largely unaffected by
non-magnetic disorder.  In contrast, $\rho(E) \propto |E|$ in clean
d-wave superconductors, and can be substantially altered by disorder.
Much of the current understanding of disorder effects comes from
perturbative theories, such as the widely-used self-consistent
T-matrix approximation (SCTMA).  In particular, the SCTMA is exact in
the limit of a single impurity, and has been used in studies of the
local DOS near an isolated scatterer\cite{Byers}.  For sufficiently
strong scatterers, an isolated impurity introduces a pair of
resonances at energies $\pm \Omega_0$\cite{Salkola,Fehrenbacher},
where $\Omega_0 < \Delta$ is a function of 
of the impurity potential $u_0$ and of the band
asymmetry.  Analytic expressions for $\Omega_0(u_0)$ have been given
for a symmetric band\cite{Salkola}, and in this special instance the
unitary limit $\Omega_0 \rightarrow 0$ coincides with $u_0 \rightarrow
\infty$.  For a realistic (asymmetric) band, the relationship is more
complex\cite{Fehrenbacher}.

For a finite concentration of impurities $n_i$, the SCTMA predicts that the
impurity resonances broaden, with tails which overlap at the Fermi
energy, leading to a finite residual DOS
$\rho(0)$\cite{Fehrenbacher,Gorkov,SR,Hirschfeld,Joynt,Xiang}.  The
region over which $\rho(E)\approx \rho(0)$, the
``impurity band'', and has a width comparable to the scattering rate
$\gamma$ at $E=0$.  In the Born limit, the $\pm\Omega_0$ resonances are
widely separated in energy, and the overlap of their tails is
exponentially small.  In the strong-scattering limit, however, the
overlap is substantial and $\gamma \sim \sqrt{\Gamma \Delta_d}$, where
$\Delta_d$ is the magnitude of the d-wave gap, $\Gamma = n_i/\pi N_0$
is the scattering rate in the normal state, and $N_0$ is the 2D
normal-state DOS at the Fermi level.  Several recent
experiments\cite{Ishida,Bonn,Taillefer,Kapitulnik} have studied
quasiparticles in the impurity band in Zn-doped high $T_c$ materials.
Of particular note are attempts to verify a provocative
prediction\cite{Bonn,Taillefer} that transport coefficients on energy
scales $\omega, \, T < \gamma$ have a universal value, independent of
$\Gamma$\cite{Lee,Scalapino}.

The SCTMA which forms the basis of this world-view has several
limitations.  It is an effective medium theory, in which one solves
for the eigenstates of an isolated impurity in the presence of a
homogeneous mean-field representing all other impurities.  This
approach ignores multiple impurity scattering processes which are
responsible for localization physics in metals and may lead to novel
effects in 2D d-wave superconductors\cite{Lee,Nersesyan,Fisher}.
Another limitation of most SCTMA calculations is the use of a
$\delta$-function potential as an impurity model.  While this
simplifies the calculation substantially, numerical results hint that
the detailed structure of the impurity potential may be
important\cite{Xiang}.  A related issue, which will be discussed at
length in this Letter, is inhomogeneous order-parameter suppression.
It is well-known that d-wave superconductivity is destroyed locally
near a strong scatterer, and in the single-impurity
limit\cite{Shnirman,atkinson}, the additional scattering was found to
renormalise $\Omega_0$ but, surprisingly, to leave most other details
of the scattered eigenstates unchanged.  Here we show, using exact
numerical solutions of the Bogoliubov-deGennes (BdG) equations, that
additional novel physics {\em does} arise in the many-impurity case.

\begin{figure}
\begin{center}
\leavevmode
\epsfxsize 0.8\columnwidth
\epsffile{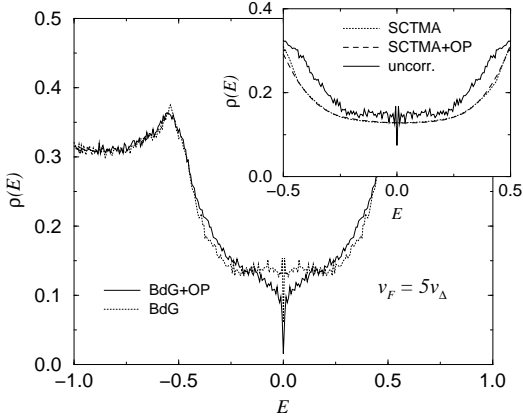}
\caption{Density of states for $n_i = 0.04$ and $u_0 = 10$.  Numerical
solutions of the BdG equations are shown with (BdG+OP) and without
(BdG) self-consistent calculation of $\Delta_{ij}$. Inset: T-matrix
calculations of $\rho(E)$ with (SCTMA+OP) and without (SCTMA)
off-diagonal scattering.  Also shown is a model with uncorrelated
impurity and off-diagonal potentials (uncorr.).}
\label{fig1}
\end{center}
\end{figure}

The main results of this work are summarised in Fig.~\ref{fig1}.
For a homogeneous order parameter, $\rho(E)$ saturates
at a constant value as $E\rightarrow 0$ (in agreement with SCTMA), 
down to a mesoscopic energy
scale $\sim 1/\rho L^2$ where level repulsion across the Fermi surface
induces a gap.  This small gap may be a precursor to a
regime associated with strong localization in the thermodynamic limit as
discussed by Senthil et al.\cite{Fisher}.  These authors predicted
asymptotic power-law behavior in $\rho(E)$ over an exponentially small
energy scale $E_2 \sim 1/\rho(0)\xi_L^2$, where $\rho(0)$ is the
residual density of states in the impurity band plateau and $\xi_L \sim
(v_F/\gamma) \exp(v_F/v_\Delta + v_\Delta/v_F)$ is the quasiparticle
localization length.  ($v_F$ is the Fermi velocity and $v_\Delta$ is
the gradient of $\Delta_k$ along the Fermi surface at the gap node).
The actual value of the DOS at $E=0$ is consistent with zero but not
determined in our work; theoretically this point is still
controversial.\cite{Ziegler}

\begin{figure}[t]
\begin{center}
\leavevmode
\epsfxsize \columnwidth
\epsffile{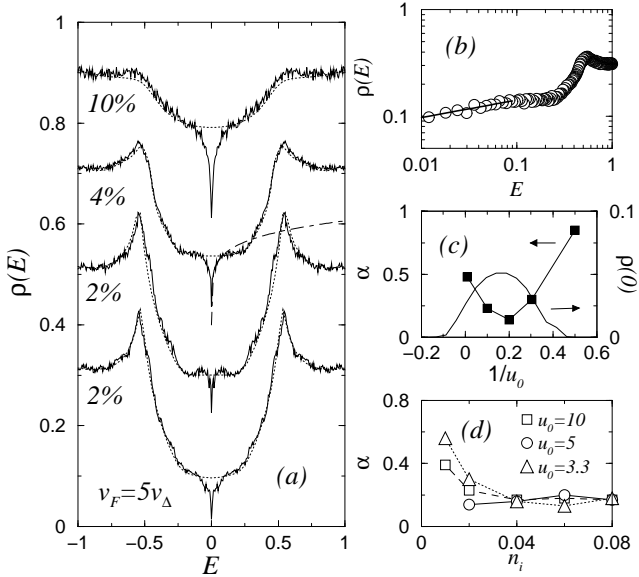}
\caption{Dependence of $\rho(E)$ on $n_i$ and $u_0$: (a) BdG+OP (solid)
and SCTMA (dotted) for $u_0 = 5$ (top three curves) and
$u_0=10$ (bottom).  Also shown (dot-dash) is a fit $\rho(E) =
A|E|^\alpha$ for $n_i = 0.04$.  (b) Logarithmic plot showing
power law behavior for $n_i=0.04$, $u_0=5$.  (c) Dependence
of power law on $u_0$ for $n_i=0.02$.  For comparison, SCTMA
of $\rho(0)$ vs.\ $u_0$ is shown. Unitary limit is $u_0 \approx 5$. (d)
Scaling of power law with $n_i$.  Note that when $n_i=0$, $\alpha=1$.}
\label{fig2}
\end{center}
\end{figure}

\begin{figure}
\begin{center}
\leavevmode
\epsfxsize 0.8\columnwidth
\epsffile{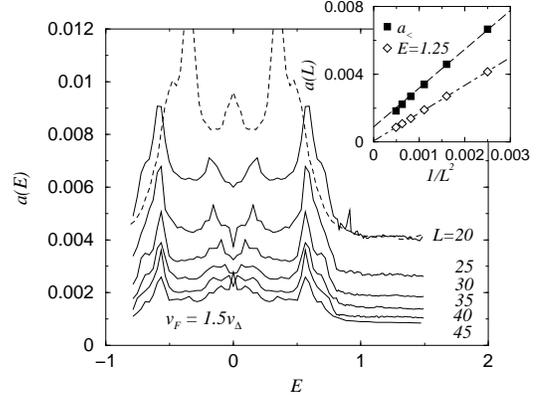}
\caption{Scaling of the inverse participation ratio for BdG+OP (solid
curves) calculations.  A BdG calculation (dashed) is also shown for
$L=20$.  Parameters are $V=-4.47$ ($\Delta_d=1.34$), $n_i=0.06$, and
$\mu=1.0$ with between 14 (L=45) and 500 (L=20) impurity
configurations and real boundary-conditions.  Inset: Scaling inside
and outside the impurity band.  $ a_< $ is the average of $a(E)$ for
$|E|<0.8$.}
\label{fig3}
\end{center}
\end{figure}

Figure~\ref{fig1} shows that when the order parameter is determined
{\it self-consistently} from the BdG equations $\rho(E)$ is quite
different.  At low energies the DOS can be fit to a power law
$\rho(E)\sim |E|^\alpha$ with nonuniversal $\alpha$
(Fig.~\ref{fig2}). The power-law is the result of spatial
correlations between the order-parameter and the impurity potential,
and is therefore fundamentally different from those of Nersesyan {\em
et al.}\cite{Nersesyan} and Senthil {\em et al}\cite{Fisher}, where
asymptotic power-laws were found, with $\alpha = 1/7$
and $\alpha = 1$ respectively.  Unlike the DOS, the dimensionless
conductance and inverse participation ratio (Fig.~\ref{fig3}) are not
changed significantly by self-consistency.  Finally, we remark that
the energy scale for the low-energy regime is $\gtsim \Omega_0$, which
is orders of magnitude larger than $E_2$ for realistic parameters.

We employ a one-band lattice model with nearest neighbour hopping
amplitude $t$ and a nearest neighbour attractive interaction $V$. 
Substitutional impurities are represented by a change in the on-site
atomic energy.  The Hamiltonian is
\begin{eqnarray}
\cal{H} &=& - t \sum_{\langle i,j\rangle} \sum_{\sigma}
c^\dagger_{i\sigma} c_{j\sigma} - \sum_{i,\sigma} [\mu - U_i]
c^\dagger_{i\sigma} c_{i\sigma} \nonumber \\ &-& \sum_{\langle
i,j\rangle} \{ \Delta_{ij} c^\dagger_{i\uparrow}
c^\dagger_{j\downarrow} + h.c. \},
\label{eq:ham}
\end{eqnarray}
where the angle-brackets indicate that site indices $i$ and $j$ are
nearest neighbours, $U_i$ is the impurity potential which takes the
value $u_0$ at a fraction $n_i$ of the sites and is zero elsewhere,
and $\Delta_{ij} = -V \langle c_{j\downarrow}c_{i\uparrow} \rangle$ is
the mean-field order parameter, determined self-consistently by
diagonalizing Eq.~(\ref{eq:ham}).  Throughout this work, energies are
measured in units of $t$, where $t$ is of order 100 meV for high $T_c$
materials.  In non-self-consistent calculations the OP has the
familiar $k$-space form $\Delta_k = \Delta_d [ \cos(k_x) - \cos(k_y)
]$, where $ \Delta_d = \frac{1}{2}\sum_\pm ( \Delta_{i\,i\pm x} -
\Delta_{i\,i\pm y} ) $ is independent of $i$.  Unless otherwise
stated, $V=-2.3$ and $\mu=1.2$ which yields $\Delta_d = 0.4$
(corresponding to $v_F/v_\Delta \approx 5$) in the absence of
disorder.  Self-consistent solutions show that $\Delta_{ij}$ is suppressed
within a few lattice constants of each strongly-scattering impurity.
Throughout this Letter, curves marked BdG and BdG+OP refer to the
neglect or inclusion of self-consistency in $\Delta_{ij}$.

The DOS is $\rho(E) = L^{-2} \sum_\alpha \delta(E-E_\alpha)$, where
$L$ is the linear system size and $E_\alpha$ are the discrete
eigenenergies of ${\cal H}$.  Our numerical calculations were
performed on periodically continued systems with $L \leq 45$.  Typical
DOS curves were obtained for $L=25$ by averaging $\rho(E)$ over $\sim
50-500$ impurity configurations and $\sim 50-100$ $k$-vectors in the
supercell Brillouin-zone.  For system sizes $L \ge 35$, computational
constraints restricted us to real periodic and anti-periodic boundary
conditions.

An important motivation for the present study is the need for a test
of the reliability of SCTMA predictions.  Thus, we use the same
disordered lattice model for the SCTMA as for the BdG calculations.  We
would also like to model order-parameter suppression within a
self-consistent T-matrix approximation (SCTMA+OP),\cite{Hettler}
and follow the ansatz of \cite{Shnirman} that the off-diagonal potential
is $\delta\Delta_{ij} = - \Delta_{ij} [ \delta_{i,0} + \delta_{j,0}
]$.  This term appears in the off-diagonal block of the effective
potential.  In both cases, the T-matrix is a $2\times 2$ matrix in
particle-hole space which satisfies
\begin{mathletters}
\begin{equation}
T_{ij}(E)= U_{ij} + \sum_{R,R^\prime} U_{iR} 
G(R-R^\prime,E) T_{R^\prime j}(E),
\label{eq:tmat}
\end{equation}
where $G$ has the Fourier transform,
\begin{equation}
G_k(E) = \big[ E \tau_0 - \epsilon_k\tau_3 -\Delta_k\tau_1
-\Sigma_k(E) \big]^{-1},
\label{eq:gfn}
\end{equation}
\label{eq:sctm}
\end{mathletters}
\noindent\noindent $\epsilon_k$ is the tight-binding dispersion,  
$\tau_i$ are the Pauli matrices, and 
$\Sigma_k(E) = n_i T_{kk}(E)$.
Equation~(\ref{eq:tmat}) is solved in real-space to take advantage of
the short range of the effective potential $U$.  Finally,
$\rho(E) = -\pi^{-1} L^{-2} \mbox{Im } \sum_k \mbox{Tr
}G_k(E)$, where the trace is over particle-hole indices.

Except for the mesoscopic gap discussed previously, 
the BdG curve in Fig.~\ref{fig1} is quantitatively similar to the SCTMA
(inset).  In contrast, the BdG+OP curve vanishes smoothly
as $E\rightarrow 0$, indicating that qualitatively new physics has
been introduced by the inclusion of order parameter suppression.  To
emphasize that this result is unexpected, we point to the popular
``Swiss cheese'' model, in which 
it is assumed that pair-breaking causes a pocket of normal metal of
radius $\xi_0$ (the coherence length) to form around each impurity.
Within this model, $\rho(0)$ is {\em enhanced} relative to the SCTMA
by $\sim n_i N_0 \xi_0^2$.

We emphasize that the correct spatial correlations between an impurity
configuration and its self-consistent off-diagonal potentials must be
preserved for the BdG+OP results to arise.  This point is illustrated
with a simple numerical calculation (inset of Fig.~\ref{fig1}), in
which $\Delta_{ij}$ is found self-consistently for random impurity
distributions which are different from those appearing in the diagonal
block of $\cal H$.  The system therefore has two distinct types of
impurity, one of which is purely off-diagonal, with uncorrelated
distributions.  It is striking that there is no hint of the correct
low-energy behaviour in this calculation.

It is instructive to compare the BdG+OP result with the SCTMA+OP
(inset, Fig.~\ref{fig1}), since they include off-diagonal scattering
from the order parameter at different levels of approximation.
Furthermore, the SCTMA+OP does preserve the correlation between
impurity location and order-parameter suppression. It has been used
succesfully to describe shifts in $\Omega_0$ due to off-diagonal
scattering\cite{Shnirman,atkinson} but, here, fails to reproduce the
correct low energy DOS in the bulk disordered case.  Instead,
$\rho(E)$ is quantitatively similar to the SCTMA, which is a direct
result of the relative smallness of the off-diagonal potential
$\Delta_d/u_0 \approx 0.04$.

In Fig.~\ref{fig2} we illustrate the dependence of the low energy DOS
on both $n_i$ and $u_0$.  In Fig.~\ref{fig2}(a) a series of curves
shows how the low-energy regime scales towards zero-width as $n_i
\rightarrow 0$ for the near-unitary scattering potential $u_0 =
5$\cite{bandasymmetry}.  That the size of the  regime should
scale faster than $\gamma$ with $n_i$ is consistent with our earlier
assertion that the novel behavior stems from a multiple-impurity
effect.  The details of the dilute impurity limit depend on the
particular value of $u_0$ however: at $2\%$ impurity concentration,
the low-energy regime is significantly larger for $u_0=10$, which lies
farther from unitarity, than for $u_0=5$.  For fixed $n_i$, we find
that the low-energy regime is $\gtsim \Omega_0$.

In Fig.~\ref{fig2}(b), a logarithmic plot of $\rho(E)$ reveals that
the low energy DOS has an apparent power-law dependence on $E$, with
non-universal exponent $0< \alpha < 1$ [Fig.~\ref{fig2}(c),(d)].  At
low impurity concentrations, $\alpha$ is a strong function of both
$n_i$ and $u_0$, with $\alpha$ a minimum for unitary scatterers.  For
larger $n_i$, $\alpha$ appears to saturate at a value which is
independent of $u_0$.  We assert that this power law is fundamentally
different from those reported elsewhere \cite{Nersesyan,Fisher}, since
it is only observed when off-diagonal scattering is present and, as we
will see next, is unrelated to strong quasiparticle localization.

We have studied the scaling of both the inverse participation ratio
$a(E)$ and the Thouless number $g(E)$ with system size.  The inverse
participation ratio is defined in the usual way, $a(E) =
x_4(E)/x_2(E)^2$, where $x_m(E) = [\rho(E)L^2]^{-1}\sum_{n,r}
[|u_n(r)|^m + |v_n(r)|^m]\delta(E-E_n)$ and $u_n(r)$ and $v_n(r)$ are
the particle and hole eigenfunctions, and is plotted in
Fig.~\ref{fig3} for several system sizes.  $a(E)$ is a direct measure
of the spatial extent of the wavefunction; it scales as $1/L^2$ for
extended states and is constant for states with localization length
$\xi_L < L$.  The scaling and energy dependence of $a(E)$ is in
general agreement with ref.~\cite{Franz} where calculations were made
with similar parameters---there is a crossover in scaling between $E <
\gamma$ and $E>\gamma$ consistent with a shorter localization length
in the impurity band.  Unlike ref.~\cite{Franz}, we do not find
saturation in $a_<$ [the average of $a(E < \gamma)$] for $L \leq 45$.
For larger disorder concentrations (not shown) where $\xi_L < 45$
sites, we find that $a(E)$ saturates at $E=0$ first indicating that
$\xi_L(E)$ is an increasing function of $E$ in the impurity band.  The
most important result for this work is that $a(E)$, and therefore
$\xi_L$, is not significantly different in the BdG and BdG+OP
calculations despite a substantial difference in $\rho(E)$.  In the
calculation which is shown, $a(E)$ is actually {\em decreased}
slightly by self-consistency, corresponding to a slight {\em increase}
in the localisation length.

We have also studied the Thouless number, defined as $g(E) = \sum_n
[E_n(\pi)-E_n(0)] \delta(E-E_n)$, where the argument of $E_n$ refers
to the application of periodic or anti-periodic 
boundary conditions in the $x$-direction.  As $L$ is
increased, $g(E)$ is expected to cross over from a constant in the
diffusive regime to exponential scaling indicative of strong
localization.  For $L \leq 45$
we find no significant scaling of $g(E)$ with $L$, consistent
with what is found for $a(E)$.  Most significantly, we find that 
$g(E)$ is nearly identical in the BdG and BdG+OP calculations, even
for low-energy states where substantial changes in $\rho(E)$ occur.

This behaviour is reminiscent of what is seen in Hartree-Fock studies
of interacting electrons in disordered conductors\cite{yang93}.
There, the Coulomb interaction enforces spatial correlations between
the disorder and charge distributions and leads to the formation of a
gap in $\rho(E)$\cite{escg}, yet leaves the dimensionless conductance
(a two-particle property related to the Thouless number) unchanged.
The power-law DOS we observe here may have a similar origin; it is
certainly clear that $\rho(E)$ depends crucially on the spatial
correlations between the impurity potential and d-wave order
parameter.  In the current work, however, it is the pairing interaction
which is relevant and the BdG equations provide a mean-field description
of the pairing interaction which is analogous to the Hartree-Fock
description of the Coulomb interaction.  We speculate that the
short-ranged pairing interaction produces spatial correlations between
distant impurities via the overlap of the long range
tails\cite{Salkola} of the single impurity resonances.

In this work we have shown that spatial correlations between order
parameter and impurity distributions in d-wave superconductors lead to
apparent power-laws in $\rho(E)$ at low energies.  These results are
potentially relevant to quasi-2D superconductors like BSCCO-2212.
Unfortunately, most disorder studies have been performed on the
anisotropic 3D YBCO system, where correlation effects are expected to
be less pronounced.  Indeed there is considerable evidence,
particularly in Zn-substituted YBCO, that disorder does indeed induce
a finite DOS at the Fermi level\cite{Ishida,Bonn,Kapitulnik} and
somewhat weaker evidence that it scales with disorder in accordance
with the SCTMA\cite{Ishida}.  Our work should therefore provide a
strong motivation to study Zn doping, and other types of planar
disorder, in the quasi-2D BSCCO-2212 system at low temperatures.

This work is supported by NSF grants  NSF-DMR-9974396 and NSF-DMR9714055.
The authors would like
to thank M.P.A. Fisher, K.\ Muttalib, S. Vishveshwara 
and K.\ Ziegler for helpful discussions.


\begin{thebibliography}{stuffstuffstuff}

\bibitem{Byers} For a recent review see M. E. Flatt\'e and
J. M. Byers, Solid State Phys. {\bf 52}, 137 (1999).

\bibitem{Salkola} A. V. Balatsky, M. I. Salkola, and A. Rosengren,
Phys. Rev. B {\bf 51} 15 547 (1995); A. V. Balatsky and
M. I. Salkola, Phys. Rev. Lett. {\bf 76}, 2386 (1996).

\bibitem{Fehrenbacher} R. Fehrenbacher, Phys. Rev. Lett. {\bf 77}, 1849
(1996);  R. Fehrenbacher and M. R. Norman, Phys. Rev. B {\bf 50} R3495
(1994);  R. Fehrenbacher, Phys. Rev. B {\bf 54}, 6632 (1996).

\bibitem{Gorkov} L.P. Gor'kov and P. A. Kaugin, JETP Lett. {\bf 41},
253 (1985).

\bibitem{SR} S. Schmitt-Rink, K. Miyake, and C. M. Varma,
Phys. Rev. Lett. {\bf 57}, 2575 (1986).

\bibitem{Hirschfeld} P. J. Hirschfeld, D. Vollhardt, and P. W\"olfle,
Solid State Commun. {\bf 59}, 111 (1986).

\bibitem{Joynt} R. Joynt, J. Low Temp. Phys. {\bf 109}, 811 (1997).

\bibitem{Xiang} T. Xiang and J. M. Wheatley, Phys. Rev. B {\bf 51},
11 721 (1995).

\bibitem{Ishida} K. Ishida {\em et al.}, J. Phys. Soc. Jpn. {\bf 62},
2803 (1993).

\bibitem{Bonn} A. Hosseini {\em et al.}, (unpublished); Kuan Zhang
{\em et al.}, Phys. Rev. Lett. {\bf 73}, 2484 (1994); S. Kamal, Ruixing Liang,
A. Hosseini, D. A. Bonn, and W. N. Hardy, Phys. Rev. B {\bf 58}, R8933 (1998).

\bibitem{Taillefer} Louis Taillefer {\em et al.}, Phys. Rev. Lett.
{\bf 79}, 483 (1997).

\bibitem{Kapitulnik} D. L. Sisson, {\em et al.}, cond-mat/9904131.

\bibitem{Lee} P. A. Lee, Phys. Rev. Lett. {\bf 71}, 1887 (1993).

\bibitem{Scalapino} P. J. Hirschfeld, W. O. Putikka, and D. J.
Scalapino, Phys. Rev. Lett. {\bf 71}, 3705 (1993);
Phys. Rev. B {\bf 50}, 10 250 (1994).

\bibitem{Nersesyan} A. A. Nersesyan, A. M. Tsvelik, and  F. Wenger,
Nucl. Phys. B {\bf 438}, 561 (1995); Phys. Rev. Lett. {\bf 72}, 2628 (1994).


\bibitem{Fisher} T. Senthil, Matthew P. A. Fisher, Leon Balents,
and Chetan Nayak, Phys. Rev. Lett. {\bf 4704}, 1998; T. Senthil
and Matthew P. A. Fisher, cond-mat/9810238.

\bibitem{Hettler} Matthias H. Hettler and P. J. Hirschfeld,
Phys. Rev. B {\bf 59}, 9606 (1999).

\bibitem{Ziegler} K. Ziegler, M.H.
Hettler, and P.J. Hirschfeld,  Phys. Rev. B 57, 10825 (1998).

\bibitem{Shnirman} Alexander Shnirman, \.{I}nanc Adagideli, Paul
M. Goldbart, and Ali Yazdani, Phys. Rev. B {\bf 60}, 7517 (1999).

\bibitem{atkinson} W. A. Atkinson, P. J. Hirschfeld, and A. H. MacDonald,
cond-mat/9912168.

\bibitem{bandasymmetry}
For $\mu=1.2$, the band is asymmetric and the unitary limit
corresponds to $u_0 \approx 6$\protect\cite{atkinson}.  Thus, $u_0=5$ is a 
{\em stronger} scatterer of low energy quasiparticles than $u_0=10$
See \protect\cite{Joynt,Fehrenbacher} for further discussion.

\bibitem{Franz} M. Franz, C. Kallin, and A. J. Berlinsky, Phys. Rev. B
{\bf 54}, R6897 (1996).

\bibitem{escg} A.L. Efros and B.I. Shklovskii, J. Phys. C {\bf 8}, L49
(1975).

\bibitem{yang93} S.-R. Eric Yang and A.H. MacDonald,
Phys. Rev. Lett. {\bf 70}, 4110 (1993).


\end{thebibliography}
\end{document}